\documentclass[aps,twocolumn,amsmath,amssymb,floatfix]{revtex4}

\usepackage{epsfig}
\usepackage{epstopdf}
\usepackage{graphicx}
\usepackage{grffile}
\usepackage{bm}
\usepackage[innercaption]{sidecap}
\sidecaptionvpos{figure}{t}
\usepackage[colorlinks,citecolor=blue,linkcolor=blue]{hyperref}
\usepackage[usenames,dvipsnames,svgnames]{xcolor}
\usepackage{array}
\usepackage{float}
\usepackage{lipsum}
\usepackage{braket}
\usepackage[normalem]{ulem} 

\newcommand{\red}[1]{\textcolor{red}{#1}}

\newcommand{\blue}[1]{\textcolor{blue}{#1}}

\renewcommand{\section}[1]{{\par\it #1---}\ignorespaces}
\renewcommand{\subsection}[1]{{\it #1 ---}\ignorespaces}

\newcommand{\vc}[1]{\mathbf{#1}}

\newcommand{\I}{i}

\begin{document}

\title{Odd-frequency superconductivity and Meissner effect in the doped topological insulator Bi$_2$Se$_3$}
\author{Johann Schmidt}
 \affiliation{Department of Physics and Astronomy, Uppsala University, Box 516, SE-751 20 Uppsala, Sweden}
\author{Fariborz Parhizgar}
\affiliation{Department of Physics and Astronomy, Uppsala University, Box 516, SE-751 20 Uppsala, Sweden}
 \author{Annica M.~Black-Schaffer}
  \affiliation{Department of Physics and Astronomy, Uppsala University, Box 516, SE-751 20 Uppsala, Sweden}
\date{\today}

\begin{abstract}
Doped Bi$_2$Se$_3$ is proposed to be a nematic superconductor originating from unusual inter-orbital pairing. We calculate all induced superconducting pair correlations in Bi$_2$Se$_3$ and discover that intra-orbital odd-frequency pairing clearly dominates over a significant range of frequencies. Moreover, we explore the contributions of even- and odd-frequency pairing to the Meissner effect, including separating intra- and inter-band processes in the response function. Contrary to expectations, and due to inter-band contributions, we find a diamagnetic Meissner effect from the odd-frequency pairing that stabilizes the superconducting order.
\end{abstract}

\maketitle

%
%
%

%
%
Superconductivity in Bi$_2$Se$_3$ has been a focal point of experimental and theoretical research in the past few years. Bi$_2$Se$_3$ had already gained prominence as a strong topological insulator \cite{Zhang2009, Xia2009}, when it was discovered that electron doping by intercalation of Cu (and later Nb and Sr) leads to the appearance of superconductivity \cite{Hor2010}. Due to the strong spin-orbit coupling, the pairing was already early on proposed to be of odd parity, making doped Bi$_2$Se$_3$ an intrinsic topological superconductor \cite{Fu2010}. More recently, a series of experiments have additionally discovered a surprising breaking of the rotation symmetry of Bi$_2$Se$_3$ in the superconducting state \cite{Matano2016, Yonezawa2016, Pan2016, Asaba2017, Shen2017, Yonezawa2018}. This so-called nematic superconducting state has been found to only be consistent with an exotic inter-orbital order parameter, meaning the Cooper pairs consist of two electrons from different orbitals in the Bi$_2$Se$_3$ low-energy structure \cite{Fu2014}, in contrast to more standard intra-orbital pairing.

In the presence of an additional electronic degree of freedom, like the orbital index in Bi$_2$Se$_3$, the symmetry classification of superconducting pairing is extended \cite{BlackSchaffer2013multi, Triola2019}. In particular, in the presence of spin, spatial parity, orbital (or band, layer, valley etc.), and time/frequency quantum numbers a total eight classes are possible, of which four are odd in frequency \cite{BlackSchaffer2013multi,Triola2019}. Odd-frequency superconductivity has the electrons paired at different times, with an odd relative time difference. It was originally envisaged as an order parameter for $^3$He \cite{Berezinskii74} and has played a major role in the understanding of proximity-induced superconductivity in heterostructures \cite{Bergeret2001,Bergeret2005,Tanaka2007,Tanaka2007b, BlackSchaffer2013TI, BlackSchaffer2013TIhetero, Linder2015, Linder2017}. More recently, bulk systems in the presence of inter-orbital terms in the Hamiltonian have also been shown to often host odd-frequency pair correlations \cite{BlackSchaffer2013multi, Parhizgar2014, Komendova2017, Kuzmanovski2017, Triola2016, Triola2018}.

As odd-frequency pairing joins electrons at different times, direct probes of this intrinsically dynamical state have proven to be challenging. Nevertheless, there still exist directly measurable consequences, such as finite Josephson currents through half metals and Weyl semimetals, entirely carried by odd-frequency pairs \cite{Eschrig2003, Asano2007,Parhizgar2018,Dutta2019}. In addition, the presence of odd-frequency correlations has been connected indirectly to for example the Kerr effect \cite{Komendova2017, Triola2018}. Another direct physical probe of odd-frequency pairing has traditionally been the Meissner effect. In contrast to the usual diamagnetic Meissner response of even-frequency pairing, which expels an external magnetic field from the superconductor, quasiclassical approaches have predicted a telltale paramagnetic Meissner effect for the proximity-induced odd-frequency pairing in heterostructures \cite{Tanaka2005, Mironov2012, Linder2017, Ouassou2019}, which was recently also experimentally observed in superconductor-ferromagnet junctions \cite{DiBernado2015}. A paramagnetic response for odd-frequency pair correlations has also been calculated for generic multiband Hamiltonians \cite{Asano2015}. Because a paramagnetic Meissner effect means the superconductor attracts instead of repels magnetic fields, the superconducting state should become unstable \cite{Abrahams1995}. While an odd-frequency order parameter has been shown to possibly be thermodynamically stable \cite{Coleman1994, Heid1995}, the paramagnetic response from odd-frequency pair correlations would prove detrimental to superconductivity in multi-orbital systems. It would therefore be very important to discover odd-frequency pairing with a diamagnetic Meissner response.

In this letter, we calculate all pair correlations induced in the inter-orbital nematic superconductor Bi$_2$Se$_3$ and discover prominent intra-orbital pairing terms, which are odd in frequency and clearly dominate the even-frequency pairing over a wide range of frequencies. Moreover, we calculate the Meissner response of both the even- and odd-frequency pair correlations and analyze inter- and intra-band processes separately. Surprisingly, we obtain a diamagnetic Meissner effect for the odd-frequency components. This is both paramount for the stability of the dominant odd-frequency pairing in the superconducting phase of Bi$_2$Se$_3$ and opens a road for designing other stable odd-frequency superconductors. Finally, we also find a clear two-fold rotational symmetry in the Meissner response signaling the nematic state. 

%
%
\section{Model} The low-energy physics in the normal state of Bi$_2$Se$_3$ is well-captured by a linear momentum model with two orbitals \cite{Fu2010,Rosenberg2012}
\begin{align}
{\cal \hat{H}}_{0}=m\sigma_x+v(k_xs_y-k_ys_x)\otimes \sigma_z+v_zk_z\sigma_y-\mu,
\label{eq:Ham}
\end{align}
where $\sigma_i$ and $s_i$ indicate the Pauli matrices in orbital and spin spaces, respectively, $v$ and $v_z$ are the Fermi velocities of the electrons in- and out-of-plane, $m$ hybridizes the different orbitals, and $\mu$ is the chemical potential. We also set $\hbar=1$. The eigenvalues of ${\cal \hat{H}}_{0}$ are given by $\epsilon^0_\pm=\pm \sqrt{m^2 + v^2(k_x^2+ k_y^2) + v_z^2k_z^2} - \mu$, forming a gapped three-dimensional (3D) Dirac dispersion, with the two-fold degenerate valence and conduction bands separated by a $2m$ energy gap.

We introduce superconductivity through the pairing matrix $\hat{\Delta}$ and write the full Hamiltonian in Nambu space as $\check{H} = \check{H}_0 + \check{\Delta}$, where $\check{H}_0=\begin{pmatrix}\hat{\mathcal{H}}_{0}({\bf k})&0\\0&- \hat{\mathcal{H}}^*_{0}(-{\bf k})\end{pmatrix}$, and $\quad \check{\Delta}=\begin{pmatrix}0&\hat{\Delta}\\ \hat{\Delta}^\dagger&0\end{pmatrix}$, where we use $\hat{...}$ ($\check{...}$) to signal $4\times4$ ($8\times8$) matrices. Out of the four possible ${\bf k}$-independent pairing symmetries identified for doped Bi$_2$Se$_3$ \cite{Fu2010}, the experimental discovery of nematic superconductivity \cite{Matano2016,Yonezawa2016} singles out an unconventional frequency independent inter-orbital spin-triplet order parameter. Specifically, this state transforms according to the 2D $E_u$ irreducible representation of the $D_3$ point group of Bi$_2$Se$_3$ and has the form $(\hat{\Delta}_x,\hat{\Delta}_y)=\Delta (s_0\otimes i\sigma_y,s_z\otimes i\sigma_y)$ \cite{Fu2014}. Below the transition temperature $T_c$, the order parameter can form any linear combination, parameterized by $\hat{\Delta}=A_x \hat{\Delta}_x + A_y \hat{\Delta}_y$ with coefficients $A_{x,y}$. The nematic state is formed by $(A_x,A_y) = (\cos(\theta),\sin(\theta))$, where $\theta$ corresponds to the in-plane angle of the nematic director. With complex coefficients $(A_x,A_y) = \frac{1}{\sqrt{2}}\left(1,\I \right)$ the order parameter instead describes a chiral superconductor.

For all numerical evaluations we set $\Delta\,=\,0.3\,\mathrm{meV}$, which is similar to values measured in scanning tunneling experiments \cite{Tao2018}. The other parameters are obtained from comparison to DFT calculations as $m\,=\,-0.28\,\mathrm{eV}$, $v\,=\,0.434\,\mathrm{eV}$, and $v_z\,=\,-0.248\,\mathrm{eV}$ \cite{Rosenberg2012}. Superconductivity in Bi$_2$Se$_3$ is observed for electron doping \cite{Lahoud2013} and we therefore study $0.27\,\mathrm{eV}<\mu<0.4\,\mathrm{eV}$, ranging from the Fermi level close to the bottom of the bulk conduction band to the doped metallic regime.

%
%
\section{Odd-frequency pairing} We first perform a complete classification of the superconducting pair correlations by studying the anomalous Green's function $\mathcal{\hat{F}}$, which is directly obtained from the Matsubara Green's function\begin{align}
\check{G}(\I \omega) = (\I \omega - \check{H})^{-1} = \begin{pmatrix}
{\cal \hat{G}}&{\cal \hat{F}}\\
\hat{ \bar{\mathcal{F}}}&\hat{ \bar{\mathcal{G}}}
\end{pmatrix}.
\label{eq:Green}
\end{align}
Here, $\mathcal{\hat{G}}$ and $\mathcal{\hat{\bar{G}}}$ are the normal Green's functions in particle and hole spaces, respectively. To obtain compact analytical expressions for all types of pair correlations, we treat, in a first step, the superconducting order parameter $\Delta$ as a small quantity relative to the other energy scales and expand the Green's function to first order in $\Delta$. The anomalous part then reduces to $\mathcal{\hat{F}}^{(1)} = \mathcal{\hat{G}}_0\hat{\Delta}\mathcal{\hat{\bar{G}}}_0$, where $\mathcal{\hat{G}}_0$ is the Green's functions of the bare Hamiltonian $\mathcal{\hat{H}}_0$.

An overview of all pairing terms in doped Bi$_2$Se$_3$ as elucidated by $\mathcal{\hat{F}}^{(1)}$ for any choice of $(A_x,A_y)$ and with a classification according to their spin, parity, orbital, and frequency symmetries is shown in Table \,\ref{tab:classification} (see Supplementary Material (SM) for the full anomalous Green's function \cite{supp}, fully consistent with Table \,\ref{tab:classification}). One of the eight different terms constitutes spin-triplet, $s$-wave, inter-orbital, even-frequency pairing, i.e.~the same symmetry as the order parameter (blue in Table \,\ref{tab:classification}). Apart from this, there is only one other $s$-wave term, i.e.~$k$-independent: a spin-triplet, intra-orbital, odd-frequency pairing (red). Thus, even though the superconducting order parameter has unconventional \emph{inter}-orbital symmetry, there exists odd-frequency spin-triplet, but otherwise conventional \emph{intra}-orbital $s$-wave pairing, explicitly induced by the hybridization between orbitals $m$. We also obtain another odd-frequency term, with spin-singlet, $p$-wave, even inter-orbital symmetry. Both of these odd-frequency (and all even-frequency) pair correlations are proportional to $A_x$ and $A_y$ in such a way, that they are present in both the nematic and chiral superconducting phases. 

\begin{table}
\setlength{\tabcolsep}{2pt}
\renewcommand{\arraystretch}{1.5}
\begin{tabular}{lcccc}
\hline
Pairing $\left(\mathcal{F}^{(1)}*D \right)$ & Spin & Parity & Orbital & Freq. \\
\hline
\hline
$ \red{\mathbf{2 i A_\pm  m \omega \Delta}}$ &  $\red{\mathbf{\uparrow \uparrow,\downarrow \downarrow}}$ & $\red{\bf{s}}$ & \red{\bf{intra}} & \red{\bf{odd}} \\
$ 2 i (A \times k) v \omega  \Delta $ &  $\uparrow \downarrow-\downarrow \uparrow$ & $p_{x,y}$ & even-inter & odd \\
\hline
$ 2 A_\pm  k_z v_z \mu \Delta$ &  $\uparrow \uparrow,\downarrow \downarrow$ & $p_z$ & intra & even \\
$ 2 \left(A \cdot k \right) v  m  \Delta$ &  $\uparrow \downarrow+\downarrow \uparrow$ & $p_{x,y}$ & intra & even \\
$ 2 \left(A \times k \right) v  k_z v_z  \Delta$ &  $\uparrow \downarrow-\downarrow \uparrow$ & $d$ & intra & even \\
\hline
$ \bf{\blue{i  A_\pm (m^2 - \mu^2 +\omega ^2) \Delta}}$ &  $\bf{\blue{\uparrow \uparrow,\downarrow \downarrow}}$ & $\bf{\blue{s}}$ & \bf{\blue{odd-inter}} & \bf{\blue{even}} \\
$ i  A_\pm k_z^2 v_z^2  \Delta$ &  $\uparrow \uparrow,\downarrow \downarrow$ & $d$ & odd-inter & even \\
$ 2  A_\pm k_z v_z \Delta$ &  $\uparrow \uparrow,\downarrow \downarrow$ & $p_z$ & even-inter & even \\
$ 2 ( A\cdot k ) v  \mu  \Delta$ &  $\uparrow \downarrow+\downarrow \uparrow$ & $p_{x,y}$ & even-inter & even \\
\hline
\hline
\end{tabular}
\caption{Symmetry classification of the pair correlations in $\hat{\mathcal{F}}^{(1)}$ for generic $(A_x,A_y)$ according to their spin, parity, orbital, and frequency symmetries. Blue term has the same symmetries as the order parameter, while the red highlights the $s$-wave intra-orbital odd-frequency pairing. Here $A_\pm = A_x \pm \I A_y$, $A\cdot k = A_xk_x +A_yk_y$, $A\times k = A_x k_y - A_y k_x$, and $D = \prod_i ((\I \omega)^2 - (\epsilon_i^0)^2)$ is the common denominator with $\epsilon^0_i$ the eigenvalues of $\mathcal{\hat{H}}_0$.  \label{tab:classification}}
\end{table}

\begin{figure}
\includegraphics[width=.48\textwidth]{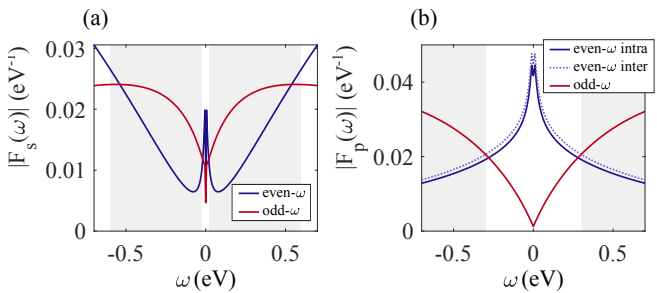}
\caption{Comparison between the absolute value $|\hat{\mathcal{F}}(\omega)|$ integrated over $\vc{k}$ for odd- and even-frequency pairing in the nematic state as a function of frequency for (a) $s$-wave and (b) in-plane $p$-wave pair amplitudes. Shaded areas mark the frequency windows at which the odd-frequency exceeds the even-frequency amplitudes obtained analytically in first order perturbation theory. Here the nematic angle is $\pi/6$. \label{fig:Fcomp}}
\end{figure}

We find that the induced odd-frequency pair amplitudes exceed the even-frequency amplitudes with the same spatial parity for a wide range of frequencies. Comparing the even- and odd-frequency $s$-wave pair amplitudes in $\hat{\mathcal{F}}^{(1)}$, the odd-frequency pairing dominates for $m+\mu < |\omega| < -m +\mu$, when the system is doped in the bulk regime, i.e.~$\mu>|m|$. The odd-frequency $p$-wave pairing becomes larger than the two even-frequency in-plane $p$-wave pair amplitudes for $|\omega| > m$, and $|\omega| > \mu$, respectively. These analytical findings are corroborated by a numerical analysis to infinite order in $\Delta$. We compare pair amplitudes as function of frequency by integrating the absolute value $|\mathcal{\hat{F}}(\I \omega)|$ over $\vc{k}$ making the usual replacement $\I \omega \rightarrow \omega \pm \I 0^+$. To study $p$-wave symmetries we multiply by the corresponding form factors before the integration. The window of frequencies in which the odd-frequency pairing dominates is clearly visible in Fig.\,\ref{fig:Fcomp} and agrees well with the analytical prediction from first order perturbation theory (shaded areas), which also further strengthens the confidence in our perturbative analysis. These results clearly demonstrate that doped Bi$_2$Se$_3$ has dominating odd-frequency superconducting pair correlations. Taking this odd-frequency pairing into account is important for any physical property sensitive to the symmetry of the superconducting pairing, such as the Josephson effect.

%
%
\section{Meissner effect} Having established the presence of dominant odd-frequency pairing in doped Bi$_2$Se$_3$, we turn to its influence on the Meissner effect. In contrast to the traditional diamagnetic Meissner effect of even-frequency superconductivity, odd-frequency pairing has traditionally been assumed to manifest itself in a paramagnetic Meissner response, which destabilizes the superconducting order \cite{Tanaka2005, Mironov2012, Linder2017, Ouassou2019, DiBernado2015, Asano2015}.

The Meissner effect is the response of a superconductor to an external magnetic field. Within linear response theory the current response $j_\mu(\vc{q},\omega_e)$ to an external vector potential $A_\mu(\vc{q},\omega_e)$ is governed by the current-current response function through $j_\mu(\vc{q},\omega_e) = -K_{\mu\nu}(\vc{q},\omega_e)A_\nu(\vc{q},\omega_e)$. Here, $\vc{q}$ and $\omega_e$ are the wave vector and angular frequency of the external vector potential, and $\mu$ and $\nu$ spatial indices $x,y,z$. The Meissner response is obtained in the limit of a static, uniform magnetic field $\omega_e \rightarrow 0, q \rightarrow 0$ (in that order) \cite{Scalapino1992, Peotta2015}. 

We introduce the vector potential in our Hamiltonian Eq.\,\eqref{eq:Ham} by replacing $\vc{k}\rightarrow\vc{k}-\vc{A}$ (setting $e = 1$) and calculate the current density operators by taking a variational derivative with respect to $A_\mu$ \cite{Liang2017}. Usually, the current density consists of a paramagnetic ($j^P \propto A_\mu^0$) and a diamagnetic part ($j^D \propto A_\mu$). However, due to the linear spectrum of the Hamiltonian $\hat{\mathcal{H}}_0$, the diamagnetic current vanishes and the current-current response function reduces to $K_{\mu\nu}(\vc{q},\omega_e) = \braket{j^P_\mu(q,\omega_e) j^P_\nu(-q,\omega_e)}$ \cite{Mizoguchi2015}, see SM for full derivation \cite{supp}. We express this expectation value to infinite order in $\Delta$ with the help of the Green's function as
\begin{align}
{K_{\mu\nu}}&= \lim_{\vc{q}\rightarrow0} \lim_{\omega_e\rightarrow0} K_{\mu\nu}(\vc{q},\omega_e) = T \sum_{\vc{k},\I \omega} \text{Tr}_e[\check{G}\check{j}^P_\mu\check{G}\check{j}^P_\nu] \nonumber\\
&= T \sum_{\vc{k},\I \omega} \text{Tr}[\mathcal{\hat{G}} \hat{j}^P_\mu\mathcal{\hat{G}}\hat{j}^P_\nu + \mathcal{\hat{F}}\hat{\bar{j}}^P_\mu\mathcal{\hat{\bar{F}}}\hat{j}^P_\nu  ], \label{eq:MeissnerGF}
\end{align}
where $\text{Tr}_e$ is a trace over only the particle part of the matrix, $T$ is the temperature, and $\hat{j}^P_\mu=-\partial{\hat{\mathcal{H}}_0(\vc{k}-\vc{A}})/\partial A_\mu$ and $\hat{\bar{j}}^P_\mu=-{\hat{j}^{P*}_\mu}$ are the paramagnetic current operators in particle and hole space, respectively. We have also suppressed the momentum and frequency dependence for legibility. 

Due to the linear dispersion of the Hamiltonian with an infinite bandwidth, the integration over the first term of Eq.\,\eqref{eq:MeissnerGF} diverges. Furthermore, it contributes an unphysical Meissner response in the limit $\Delta \to 0$, but which would be cancelled by the diamagnetic current from higher order terms of a more material-specific Hamiltonian \cite{Mizoguchi2015}. However, because the even- and odd-frequency pair correlations are naturally fully contained in the anomalous Green's function $\hat{\mathcal{F}} = \hat{\mathcal{F}}^e+\hat{\mathcal{F}}^o$, we can work out their complete influence on the Meissner response by focusing solely on the last term in Eq.\,\eqref{eq:MeissnerGF} \cite{Asano2015}. Additionally, in the limit of small $\Delta$, applicable for Bi$_2$Se$_3$, the last term in Eq.\,\eqref{eq:MeissnerGF}, even captures the full Meissner effect, because the diverging first term is then always cancelled by the the normal state contribution of $K^{(0)}_{\mu\nu}=\text{Tr}[\mathcal{\hat{G}}^0 \hat{j}^P_\mu\mathcal{\hat{G}}^0\hat{j}^P_\nu]$ \cite{Mizoguchi2015}, as we also numerically confirm, see SM \cite{supp}. 

Focusing on the contribution of the second term in Eq.~\eqref{eq:MeissnerGF}, ${K^{(S)}_{\mu\nu}}$, we find a sum of even- and odd-frequency contributions ${K^{(S)}_{\mu\nu}} = K^{e}_{\mu\nu} + K^{o}_{\mu\nu} = T\sum_{\vc{k}, \I \omega}\text{Tr}[
\mathcal{\hat{F}}^e\hat{\bar{j}}_\mu\hat{\bar{\mathcal{F}}}^e\hat{j}_\nu + \mathcal{\hat{F}}^o\hat{\bar{j}}_\mu\hat{\bar{\mathcal{F}}}^o\hat{j}_\nu]$, since the terms $\text{Tr}[{\cal \hat{F}}^e\hat{\bar{j}}_\mu \hat{\bar{{\cal F}}}^o\hat{j}_\nu],\text{Tr}[{\cal \hat{F}}^o\hat{\bar{j}}_\mu \hat{\bar{{\cal F}}}^e\hat{j}_\nu]$ vanish. The sign of ${K^{(S)}_{\mu\nu}}$ determines whether the Meissner response is dia- (${K^{(S)}_{\mu\nu}}>0$) or paramagnetic (${K^{(S)}_{\mu\nu}}<0$) \cite{Asano2015}. To further simplify the results, we use the bands obtained from diagonalizing $\check{H}$ to split the response function ${K^{(S)}_{\mu\nu}}$ into intra- and inter-band contributions. The intra-band contributions are dominated by quasiparticles excited just above the superconducting gap, while the inter-band contribution, on the other hand, becomes enhanced when two bands approach each other in the vicinity of, but not necessarily at, the Fermi surface. After carrying out the Matsubara summation analytically, we integrate numerically over $\vc{k}$ at $T=2\,\mathrm{K}$, see SM \cite{supp} for further details.

\begin{figure}[tb]
    \centering
    \includegraphics[width=.48\textwidth]{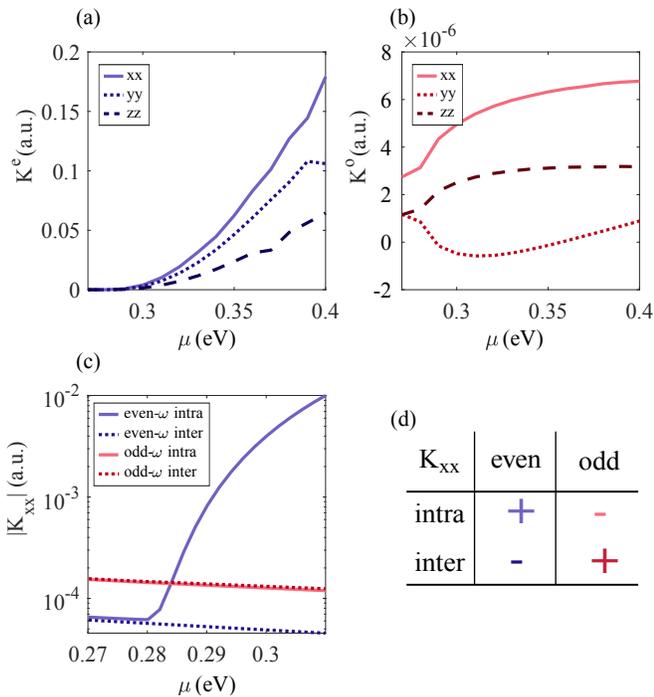}
    \caption{Superconducting contributions to the Meissner effect for the nematic state as a function of chemical potential. (a) and (b) $K^{(S)}_{xx}$, $K^{(S)}_{yy}$, and $K^{(S)}_{zz}$ responses for even- and odd-frequency contributions, respectively. (c) Intra- and inter-band parts of the $K^{(S)}_{xx}$ Meissner response for the even- and odd-frequency contributions around the onset of the conduction band. (d) Summary of signs ($+$ for diamagnetic, $-$ for paramagnetic) for the even- and odd-frequency, intra- and inter-band contributions to $K^{(S)}_{xx}$ displayed in (c). Here the nematic angle is $\pi/6$.}
    \label{fig:total}
\end{figure}

The resulting even- and odd-frequency contributions to the Meissner effect in the nematic state are presented in Figs.\,\ref{fig:total} (a,b) as a function of chemical potential. For the even-frequency contribution we find a standard diamagnetic Meissner response, which is also the dominant contribution to the total Meissner response, especially for $\mu \gg m$. For the odd-frequency contribution we find that it also contributes a \emph{dia}magnetic and not paramagnetic Meissner response, for almost all field directions and parameters. Considering that odd-frequency pairing has been widely considered to yield a paramagnetic response \cite{Tanaka2005, Mironov2012, Asano2015, Linder2017, Ouassou2019, DiBernado2015}, this is a surprising finding, which we can understand by separately studying the intra- and inter-band processes. In Figs.\,\ref{fig:total}\,(c,d) we display separately the intra- and inter-band contributions to the $K^{(S)}_{xx}$ component at doping levels around the onset of the conduction band. 
For intra-band processes, the even-frequency pairing gives rise to a diamagnetic and the odd-frequency pairing to a paramagnetic Meissner response, as expected. The inter-band processes, however, contribute with the opposite sign.
The total response can then be para- or diamagnetic depending on the balance between intra- and inter-band processes. The even-frequency pairing yields a dominating diamagnetic Meissner response, because the intra-band contribution is much larger than the inter-band one for $\mu > |m|$, as clearly seen in Fig.\,\ref{fig:total}\,(c).
For the odd-frequency pairing in doped Bi$_2$Se$_3$ we find intra- and inter-band processes of the same order of magnitude and also notably larger than the corresponding even-frequency processes below the onset of the conduction band. For the realistic parameters used in Fig.~\ref{fig:total}\,(c), the odd-frequency contributions can be strongly diamagnetic, when the inter-band processes become dominant, but also slightly paramagnetic such as for the $K^{(S)}_{yy}$ component in Fig.\,\ref{fig:total}\,(b).
As a consequence, the dominating odd-frequency pairing in doped Bi$_2$Se$_3$ actually contributes to a stabilizing diamagnetic Meissner effect. This is at odds with the previous notion of a paramagnetic and thus destabilizing Meissner response from odd-frequency pairing and also opens the possibility of finding other stable multi-orbital odd-frequency superconductors.

%
\section{Nematicity} Finally, we discuss the nematicity of the Meissner response, which is already apparent in Fig.\,\ref{fig:total}, where $K^{(S)}_{xx}>K^{(S)}_{yy}>K^{(S)}_{zz}$. A different response in the $z$-direction is expected from the different Fermi velocities, $|v|>|v_z|$. The in-plane Fermi surface is, however, isotropic, such that the observed nematicity in the in-plane Meissner response, with $K^{(S)}_{xx}$ almost twice as large as $K^{(S)}_{yy}$ for $\theta =\pi/6$ in  Fig.\,\ref{fig:total}, constitutes a clear imprint of nematic superconductivity.

\begin{figure}[tb]
    \centering
    \includegraphics[width=.48\textwidth]{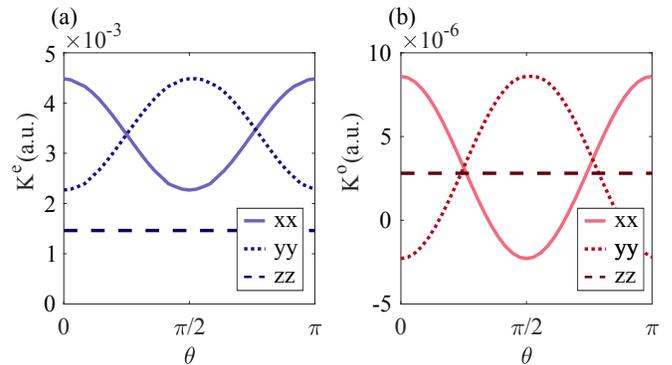}
    \caption{In- and out-of-plane (a) even- and (b) odd-frequency contributions to the Meissner response as a function of nematic angle $\theta$ for chemical potential $\mu = 0.3\,$eV. }
    \label{fig:angle}
\end{figure}
The nematicity also appears in the response for different nematic angles $\theta$ as we display in Fig.\,\ref{fig:angle}, which is equivalent to the Meissner response for a fixed nematic angle but with the magnetic field rotated in the $x$-$y$ plane. The $K^{(S)}_{xx}$ and $K^{(S)}_{yy}$ response display a clear two-fold periodicity as a function of nematic angle both for the even- and the odd-frequency response, while the out-of-plane $K^{(S)}_{zz}$ response is completely independent of the choice of $\theta$. This two-fold periodic response is a clearcut manifestation of the spontaneous rotational symmetry breaking by the superconducting order parameter, as the Meissner effect explicitly measures the superconducting state. The variation is significant, with $\text{max}(K^{(S)}_{xx})/\text{min}(K^{(S)}_{xx})\approx 2$, even though the very small order parameter hardly affects the band dispersion. 

Before concluding, we note that the response presented in Fig.\,\ref{fig:angle} only contains the superconducting contribution to the Meissner effect. The total Meissner response $K_{\mu\nu}$ requires also including the first term in Eq.\,\eqref{eq:MeissnerGF}, as well as the contribution from the diamagnetic current operator for corrections to the low-energy spectrum beyond  linear order in $\vc{k}$. However, adding any of these neglected terms will only overlay additional two- or six-fold symmetric quantities on top of the very pronounced two-fold rotation symmetry in Fig.\,\ref{fig:angle}. Thus, we still expect a clear signature of nematic superconductivity in the Meissner effect, experimentally detectable e.g.~in the London penetration depth.

%
\section{Conclusions \label{sec:conclusion}} 
In conclusion, we identify large and dominating odd-frequency intra-orbital $s$-wave pair correlations in the nematic superconducting state in doped Bi$_2$Se$_3$, even though the order parameter itself constitutes only even-frequency inter-orbital $s$-wave pairing. With odd-frequency pair correlations known to influence measurable quantities such as Josephson currents, this result is important for understanding the physical behavior of doped Bi$_2$Se$_3$. We further discover that the odd-frequency intra- and inter-band contributions to the Meissner response have opposite signs, canceling in an intricate way to give rise to a small, but unexpectedly diamagnetic Meissner response, which stabilizes the superconducting state. This changes the current understanding, where odd-frequency pairing has been assumed to give a paramagnetic Meissner signal, and reveals the possibility to engineer odd-frequency superconductors where enhanced inter-band contributions give a stable diamagnetic Meissner response.


We are grateful to D.~Kuzmanovski for helpful discussions about the Meissner kernel. This work was supported by the Swedish Research Council (Vetenskapsr\aa det Grant No. 2018-03488), the European Research Council (ERC) under the European Unions Horizon 2020 research and innovation programme (ERC-2017-StG-757553), the Swedish Foundation for Strategic Research (SSF), and the Knut and Alice Wallenberg Foundation.

\bibliographystyle{apsrevmy}

\end{document}


\title{Supplementary Material to ``Odd-frequency superconductivity and the Meissner effect in doped Bi$_2$Se$_3$"}
\author{Johann Schmidt}
 \affiliation{Department of Physics and Astronomy, Uppsala University, Box 516, SE-751 20 Uppsala, Sweden}
\author{Fariborz Parhizgar}
\affiliation{Department of Physics and Astronomy, Uppsala University, Box 516, SE-751 20 Uppsala, Sweden}
 \author{Annica M.~Black-Schaffer}
  \affiliation{Department of Physics and Astronomy, Uppsala University, Box 516, SE-751 20 Uppsala, Sweden}
\date{\today}

\maketitle

In this supplementary material we first present the complete form of the anomalous Green's function. Then we explain the derivation of the Meissner kernel in more detail and finally describe how we perform the Matsubara summation with focus on the splitting into the intra- and inter-band contributions.

%
%
\section{Anomalous Green's function to infinite order in $\Delta$\label{sec:GFSM}}

In the main text, we perform a perturbation expansion of the anomalous Green's function up to first order in the order parameter $\Delta$, with all induced pair correlations are presented in Table I of the main text. Here we report the anomalous Green's function obtained completely non-perturbatively from Eq.~2 in the main text for the nematic state. We express the anomalous Green's function in the basis $\left(1\uparrow, 1\downarrow, 2\uparrow, 2\downarrow \right)$, where $1,2$ indicate the different $p_{1z},p_{2z}$-orbitals that form the basis of the Hamiltonian of the normal-state doped Bi$_2$Se$_3$ (Eq.~1 in the main text) and $\uparrow,\downarrow$ show different spins \cite{Fu2010}. We also separate the odd- and even-frequency parts $\mathcal{F} = \mathcal{F}^o + \mathcal{F}^e$, which take the form
\begin{align}
{\cal F}^o  = \frac{2\omega\Delta_0}{D_-D_+}  \begin{pmatrix}
A_-m& 0 & 0 & -A\times k \\
0& -A_+m&A\times k & 0 \\
0 &-A\times k &-A_-m & 0 \\
A\times k& 0 & 0 & A_+m 
\end{pmatrix},
\label{FoSM}
\end{align}
and
\begin{widetext}
\begin{align}
{\cal F}^e  = \frac{2\Delta_0}{D_-D_+}  \begin{pmatrix}
A_-k_z\mu& mA\cdot k+k_zA\times k & \frac{i}{2}(A_-\gamma_--A_+k_-^2) & \mu A\cdot k \\
mA\cdot k-k_zA\times k& -A_+k_z\mu&\mu A\cdot k & -\frac{i}{2}(A_+\gamma_--A_-k_+^2) \\
-\frac{i}{2}(A_-\gamma_+-A_+k_-^2) &\mu A\cdot k &A_-k_z\mu & mA\cdot k-k_zA\times k \\
\mu A\cdot k& \frac{i}{2}(A_+\gamma_+-A_-k_+^2) & mA\cdot k + k_zA\times k & -A_+k_z\mu 
\end{pmatrix},
\label{FeSM}
\end{align}
\end{widetext}
with the shorthand notation $k_{x,y} \equiv vk_{x,y}$, $k_z \equiv v_z k_z$, $\gamma_\pm=((m\pm ik_z)^2-\mu^2-\omega^2-\Delta^2)$, $A_\pm = A_x \pm \I A_y$, $A\times k = A_x k_y - A_y k_x$, and $A\cdot k = A_xk_x +A_yk_y$. The denominator terms are given by $D_\pm = ((\I \omega)^2 - \xi_\pm^2)$, where $\xi_\pm = \sqrt{\epsilon ^2 + \Delta^2 + \mu^2 \pm \sqrt{\Delta^2(m^2 + (A\times k)^2) + \epsilon^2\mu^2}}$ are the eigenvalues of the full Hamiltonian $\check{H}$. Here $\epsilon=|\epsilon_\pm^0+\mu|$ where, $\epsilon_\pm^0$ are the eigenvalues of the normal Hamiltonian, $\epsilon_\pm^0=\pm\sqrt{m^2+k_x^2+k_y^2+k_z^2}-\mu$. It should be noted that the denominator is even in frequency and does not influence the symmetry classification.

Analyzing the non-perturbative results in Eqs.~\eqref{FoSM}-\eqref{FeSM}, we find that they are slightly more complicated but fully consistent with the results in the main text calculated perturbatively to first order in $\Delta$. The first-order results are obtained by replacing $\gamma_\pm \rightarrow ((m\pm ik_z)^2-\mu^2-\omega^2)$ and $\xi_\pm \rightarrow \epsilon_\pm^0$, which are notably also valid for the chiral state, not just the real combinations of $A_{x,y}$ which generate the nematic state.

%
%
\section{Detailed derivation of the Meissner effect}

In this section we present a more detailed derivation of the Meissner effect for doped Bi$_2$Se$_3$. The Meissner effect describes the response of a superconductor to an external magnetic field. It can phenomenologically be derived from the London equation, which relates the superconducting current density $\vc{j}$ to an external vector potential $\vc{A}$ by $
    \vc{j} = - \frac{n_s e^2}{m}\vc{A}.
$, where $n_s$ is the superconducting density, and $m$ and $e$ are the electron mass and charge, respectively. Together with the definition of the vector potential $\nabla \times \vc{A} = \vc{B}$ and Ampere's law $\nabla \times \vc{B} = \mu_0 \vc{j}$, it can be used to derive the differential equation for the magnetic field inside a superconductor:
$
    \nabla^2 \vc{B} = \frac{1}{\lambda^2}\vc{B}.
$
This differential equation yields an exponentially decaying solution and thus the magnetic field is expelled in the interior of a superconductor. Here, $\lambda = \sqrt{\frac{m}{n_s e^2 \mu_0}}$ is the London penetration depth, the characteristic length scale at which the magnetic field is suppressed. 

In the London picture, the suppression of the magnetic field occurs because the coefficient $\frac{n_s e^2}{m}$ is assumed to always be positive. Within a microscopic theory the phenomenological London equation is replaced by the more general relation presented in the main text,
\begin{align}
j_\mu(\vc{q},\omega_e) = -K_{\mu\nu}(\vc{q},\omega_e)A_\nu(\vc{q},\omega_e),
\end{align}
where $\vc{q}$ and $\omega_e$ are the wave vector and angular frequency of the external vector potential, and $\mu$ and $\nu$ represent the spatial indices $x,y,z$. The current-current correlation function $K_{\mu\nu}$ can in principle take any sign, such that the magnetic field can either be suppressed (for $K>0$) or increased ($K<0$) in the superconductor. The traditional Meissner effect, suppressing the magnetic field, is then called the diamagnetic Meissner effect in contrast to a paramagnetic Meissner effect.

Within linear response theory, we can treat the vector potential as a perturbation such that $K_{\mu\nu}(\vc{q},\omega_e)$ takes the general form \cite{Mahan, Kuzmanovski2012, Liang2017}
\begin{align}
   K_{\mu\nu} (\vc{q}, \omega_e) &= \braket{j^P_\mu(\vc{q},\omega_e) j^P_\nu(-\vc{q},\omega_e)} + \braket{j^D_{\mu,\nu}(\vc{q},\omega_e)},
\end{align}
where all expectation values are taken with respect to the unperturbed Hamiltonian and $j^P$ and $j^D$ are the paramagnetic and diamagnetic current operators, respectively. To obtain the current operators, we introduce the vector potential $\vc{A}$ in the Hamiltonian through the substitution $\vc{k} \rightarrow \vc{k} - \vc{A}$, expand to second order in $\vc{A}$, and differentiate with respect to $\vc{A}$ \cite{Scalapino1992, Liang2017}. The resulting current can be split into the sum 
$
    j^P_\mu + j^D_{\mu,\nu} A_\nu.
$
The Hamiltonian for Bi$_2$Se$_3$ presented in Eq.~(1) in the main text is linear in the crystal momentum $\vc{k}$. Thus, when we introduce the vector potential through the substitution $\vc{k}\rightarrow\vc{k}-\vc{A}$, there is no quadratic term in $\vc{A}$, and the hence $j^D_\mu$ automatically vanishes. As a result, the current-current response function in Eq.\,\eqref{eq:Kfull} reduces to the expectation value
$
K_{\mu\nu} (\vc{q}, \omega_e) = \braket{j^P_\mu(\vc{q},\omega_e) j^P_\nu(-\vc{q},\omega_e)}.
$

The Meissner effect is the response to a static, uniform external vector potential, corresponding to the limit $\omega_e \rightarrow 0, \vc{q} \rightarrow 0$. Then, expressing the expectation value of the second quantized current operators in the expression for $K_{\mu\nu}$ with the help of the Green's functions, we arrive at Eq.\,(3) in the main text:
\begin{align}\label{eq:Kfull}
{K_{\mu\nu}}&= \lim_{\vc{q}\rightarrow0} \lim_{\omega_e\rightarrow0} K_{\mu\nu}(\vc{q},\omega_e) = T \sum_{\vc{k},\I \omega} \text{Tr}_e[\check{G}\check{j}^P_\mu\check{G}\check{j}^P_\nu] = T \sum_{\vc{k},\I \omega} \text{Tr}[\mathcal{\hat{G}} \hat{j}^P_\mu\mathcal{\hat{G}}\hat{j}^P_\nu + \mathcal{\hat{F}}\hat{\bar{j}}^P_\mu\mathcal{\hat{\bar{F}}}\hat{j}^P_\nu  ].
\end{align}
Here $\check{G}$ is the full $8\times8$ Green's function, $\hat{\mathcal{G}}$ and $\hat{\mathcal{F}}$ are the $4\times4$  normal and anomalous Green's functions, respectively, $T$ is the temperature, and $\mathrm{Tr}_e$ is a trace over the particle part only of the full $8\times8$ matrix. Moreover, $\hat{j}^P$ are the first quantized paramagnetic current operators, which we easily obtained from 
\begin{align}
\check{j}_\mu &= \left(\begin{matrix}
\hat{j}_\mu & 0\\
0 & \hat{\bar{j}}_\mu
\end{matrix}\right) = -\frac{\delta \mathcal{\hat{H}}_{0}(\vc{A})}{\delta A_\mu},
\label{eq:currentops}
\end{align}
such that
$\hat{j}_x = -\hat{\bar{j}}^*_x = v s_y \otimes \sigma_z$, $\hat{j}_y = -\hat{\bar{j}}^*_y= -v s_x \otimes \sigma_z$, and $\hat{j}_z = -\hat{\bar{j}}^*_z= v_z \sigma_y$.

The vanishing diamagnetic current due to the linear dispersion also means that the normal Green's function contribution $Tr[\mathcal{G} \hat{j}_\mu\mathcal{G}\hat{j}_\nu]$ technically contributes with a finite Meissner response even in the limit $\Delta \rightarrow 0$. Usually, this contribution is cancelled by the diamagnetic current from higher order terms in $k$ in the full band structure \cite{Mizoguchi2015}, but such terms are missing here. Because we are primarily interested in the contributions due to the odd- and even-frequency superconducting pair correlations in the anomalous Green's function, we can safely consider only the superconducting contribution to the Meissner response \cite{Asano2015}, given by the terms involving the anomalous Green's functions, 
\begin{align}
\label{eq:KS}
K^{(S)}_{\mu\nu} = T \sum_{\vc{k},\I \omega} Tr[\mathcal{\hat{F}}\hat{\bar{j}}_\mu\mathcal{\hat{\bar{F}}}\hat{j}_\nu].
\end{align}
Moreover, the normal Green's function contribution $Tr[\mathcal{G} \hat{j}_\mu\mathcal{G}\hat{j}_\nu]$ not only attains a non-zero value in the absence of superconductivity, it also diverges at high energies. It is possible to regularize it by subtracting the normal-state response $Tr[\hat{\mathcal{G}}_0 \hat{j}_\mu\hat{\mathcal{G}}_0\hat{j}_\nu]$ from Eq.\,\eqref{eq:Kfull}, which provides another way of guaranteeing a vanishing Meissner response when $\Delta = 0$ \cite{Kopnin2008, Mizoguchi2015}. 
We have numerically performed this regularization and find that, $Tr[\hat{\mathcal{G}} \hat{j}_\mu\hat{\mathcal{G}}\hat{j}_\nu]-Tr[\hat{\mathcal{G}}_0 \hat{j}_\mu\hat{\mathcal{G}}_0\hat{j}_\nu]$ vanishes indeed for $\Delta =0 $, but even for the realistic small finite value $\Delta=0.3\,\mathrm{meV}$. Hence, in this case $K^{(S)}_{\mu\nu}$ in Eq.\,\eqref{eq:KS} represents the full superconducting contribution to the Meissner response, and is the quantity used in the calculations reported in the main text. A study of the full Meissner response $K_{\mu\nu}$ would thus require a Hamiltonian beyond linear order in $\vc{k}$, which goes beyond our study.

Finally, in order to analyze the Meissner kernel we split it into a summation of the even- and odd-frequency pairing contributions:
\begin{align}\label{eq:Q2}
{K^{(S)}_{\mu\nu}}&=T\sum_{\vc{k}, \I \omega}Tr[
{\cal \hat{F}}^e\hat{\bar{j}}_\mu\hat{{\cal \bar{F}}}^e\hat{j}_\nu+{\cal \hat{F}}^e\hat{\bar{j}}_\mu \hat{\bar{{\cal F}}}^o\hat{j}_\nu+{\cal \hat{F}}^o\hat{\bar{j}}_\mu \hat{\bar{{\cal F}}}^e\hat{j}_\nu+{\cal \hat{F}}^o\hat{\bar{j}}_\mu\hat{\bar{{\cal F}}}^o\hat{j}_\nu].
\end{align}
The terms $Tr[{\cal \hat{F}}^e\hat{\bar{j}}_\mu \hat{\bar{{\cal F}}}^o\hat{j}_\mu]$, and $Tr[{\cal \hat{F}}^o\hat{\bar{j}}_\mu \hat{\bar{{\cal F}}}^e\hat{j}_\mu]$ vanish for the anomalous Green's function $\hat{F}$ presented in Eqs.\,\eqref{FoSM}-\eqref{FeSM}, and we can thus define even- and odd-frequency contributions to the Meissner response: $K^e_{\mu\nu} = T\sum_{\vc{k}, \I \omega}\text{Tr}[\mathcal{\hat{F}}^e\hat{\bar{j}}_\mu\hat{\bar{\mathcal{F}}}^e\hat{j}_\nu]$ and $K^o_{\mu\nu} = T\sum_{\vc{k}, \I \omega} \text{Tr}[\mathcal{\hat{F}}^o\hat{\bar{j}}_\mu\hat{\bar{\mathcal{F}}}^o\hat{j}_\nu]$, as reported in the main text. As an example of the resulting analytical expressions, we include here the odd-frequency contribution to the Meissner response $K^{(S)}_{xx}$ in the nematic state to infinite order in $\Delta$:
\begin{align}
K_{xx}^o &= T \sum_{\vc{k}, \I \omega} \frac{8 \Delta_0^2 \omega^2}{D_-^2 D_+^2} \left( 2 (A\times k)^2 + (A_+^2 + A_-^2) m^2 \right)
\end{align}
All other responses can similarly be derived.

\section{Matsubara summation\label{sec:Matsubara}}

In this subsection we present more details on the splitting of the current-current correlation function, $K_{\mu\nu}$, into intra- and inter-band processes and how we analytically apply the summation over Matsubara frequencies. After performing the matrix multiplications and taking the trace in Eq.\,\eqref{eq:Q2}, all the Meissner kernels $\mathcal{K}^{e/o} = Tr[\mathcal{\hat{F}}^{e/o}\hat{\bar{j}}_\mu\hat{\bar{\mathcal{F}}}^{e/o}\hat{j}_\mu]$ have the general form
\begin{align}
{\cal K}=\frac{a+b(\I\omega)^2+c(\I\omega)^4}{(\I\omega-\xi_+)^2(\I\omega+\xi_+)^2(\I\omega-\xi_-)^2(\I\omega+\xi_-)^2},
\end{align} 
where $\xi_\pm$ are the eigenstates to the full Hamiltonian, given above in the Section\,\ref{sec:GFSM}.
where $a,b$, and $c$ are coefficients, different for the even- or odd-frequency contributions. In particular, these terms are dependent on the parameters of the model and are also functions of the wave vectors $k_x,k_y,k_z$, but not frequency $\I\omega$.
Following a standard approach to response functions, we split the expression for the Meissner kernel into intra- and inter-band processes. The intra-band processes are separately proportional to electronic structures of each band (not orbital) and hence they can be well-described by the quasiparticles closest to the superconducting gap. On the other hand, the inter-band processes contribute with multiple bands in the response function. In gapped systems, these processes become more important in dynamical effects and hence odd-frequency pairing, with its special frequency dependency, may experience particularly large contributions from these processes. By using this intra-/inter-band separation, the Meissner kernel splits as
\begin{align}
\mathcal{K} = \mathcal{K}^{intra} +\mathcal{K}^{inter} = \left(\frac{\alpha}{\left(\I \omega)^2- \xi_+^2\right)^2}+\frac{\beta}{\left(\I \omega)^2- \xi_-^2\right)^2}\right) + \left(\frac{\gamma}{\left(\I \omega)^2- \xi_+^2\right)\left(\I \omega)^2- \xi_-^2\right)}\right),\label{eq:split}
\end{align}
where the first parenthesis comprises the intra- and the last the inter-band processes. The coefficients $\alpha$, $\beta$, and $\gamma$ can be expressed in terms of $a$, $b$ and $c$ as
\begin{align}
\alpha &= \frac{a + b \xi_+^2 + c \xi_+^4}{(\xi_+^2 - \xi_-^2)^2}, & 
\beta &= \frac{a + b \xi_-^2 + c \xi_-^4}{(\xi_+^2 - \xi_-^2)^2}, &
\gamma &= -\frac{2a + b (\xi_+^2 + \xi_-^2) + 2c (\xi_+^2 \xi_-^2)}{(\xi_+^2 - \xi_-^2)^2}.\label{eq:coeff}
\end{align}
We then perform the Matsubara summation by using the following identities
\begin{align}
T \sum_{\I \omega} \frac{1}{\left((\I \omega)^2- \xi^2\right)^2} &= \frac{1}{2 \xi^2} (\mathcal{C}(\xi)+n^\prime(\xi)), &
T \sum_{\I \omega} \frac{1}{\left((\I \omega)^2- \xi_1^2\right)\left((\I \omega)^2- \xi_2^2\right)} = -\frac{\mathcal{C}(\xi_1) - \mathcal{C}(\xi_2)}{\xi_1^2 - \xi_2^2},\label{eq:identities}
\end{align}
where $n(\xi)$ is the Fermi-Dirac distribution function and $n'(\xi)$ is its derivative, which together with ${\cal C}(\xi)=(n(-\xi)-n(\xi))/2\xi$ are given by
$n(\xi)=\frac{1}{2}\left(1-\tanh\left(\frac{\beta\xi}{2}\right)\right)$, 
$n'(\xi)=-\frac{\beta}{4}\mathrm{sech}\left(\frac{\beta\xi}{2}\right)$, and
${\cal C}(\xi)=\frac{1}{2\xi}\tanh\left(\frac{\beta\xi}{2}\right)$.
Finally, combining Eqs.\,\eqref{eq:split}-\eqref{eq:identities} gives the Meissner kernels
\begin{align}
\mathcal{K}^{intra} &= \frac{a + b \xi_+^2 + c \xi_+^4}{2 (\xi_+^2 - \xi_-^2)^2 \xi_+^2} (\mathcal{C}(\xi_+)+n^\prime(\xi_+)) + \frac{a + b \xi_-^2 + c \xi_-^4}{2(\xi_+^2 - \xi_-^2)^2 \xi_-^2} (\mathcal{C}(\xi_-)+n^\prime(\xi_-))\nonumber\\
\mathcal{K}^{inter} &= \frac{2a + b (\xi_+^2 + \xi_-^2) + 2c (\xi_+^2 \xi_-^2)}{(\xi_+^2 - \xi_-^2)^3} (\mathcal{C}(\xi_+) - \mathcal{C}(\xi_-)).\label{eq:Kernels}
\end{align}
The superconducting contribution to the Meissner response $K^{(S)}_{\mu\nu}$, presented in Figs.~2 and 3 in the main text, are obtained by numerically integrating Eqs.\,\eqref{eq:Kernels} over $\vc{k}$.

\bibliographystyle{apsrevmy}